\documentclass[pra,twocolumn,showpacs]{revtex4}
\usepackage{epsfig}
\usepackage{graphicx}
\usepackage{dcolumn}
\usepackage{amsmath,amssymb}
\usepackage{pstricks}
\usepackage{color}
\usepackage{footnote}
\usepackage{bm}
\usepackage{bbold}
\usepackage[colorlinks=true,citecolor={blue},urlcolor={blue}, linkcolor=blue]{hyperref}
\begin{document}
%%%%%%%%%%%%%%%%%%%%%%%%%%%%%%%%%%%%%%%%%%%%%%%%%%%%%%%%%%%%%%%%%%%%%%%%%%%%%%%%%%%%%%%%%%%%%%%%%%%%%
\title{Relativistic coupled-cluster investigation of parity (${\mathcal{P}}$) and time-reversal (${\mathcal{T}}$) symmetry violations in HgF} 
%${\mathcal{P}}$,${\mathcal{T}}$-odd interactions in HgF}
%%%%%%%%%%%%%%%%%%%%%%%%%%%%%%%%%%%%%%%%%%%%%%%%%%%%%%%%%%%%%%%%%%%%%%%%%%%%%%%%%%%%%%%%%%%%%%%%%%%%%
\author{Kaushik Talukdar,$^{1,}$\footnote{talukdar.kaushik7970@gmail.com}
%Sudip Sasmal,$^{2}$ %\footnote{sudipsasmal.chem@gmail.com}
Malaya K. Nayak,$^{2,}$\footnote{mknayak@barc.gov.in, mk.nayak72@gmail.com}
Nayana Vaval,$^{3,}$\footnote{np.vaval@gmail.com}
Sourav Pal$^{4,1,}$\footnote{spal@chem.iitb.ac.in}}
%%%%%%%%%%%%%%%%%%%%%%%%%%%%%%%%%%%%%%%%%%%%%%%%%%%%%%%%%%%%%%%%%%%%%%%%%%%%%%%%%%%%%%%%%%%%%%%%%%%%%
\affiliation{$^1$Department of Chemistry, Indian Institute of Technology Bombay, Powai, Mumbai 400076,  India}
%\affiliation{$^2$Department of Physics and Astronomy, Aarhus University, Ny Munkegade 120, 8000 Aarhus C, Denmark}
\affiliation{$^2$Theoretical Chemistry Section, Bhabha Atomic Research Centre, Trombay, Mumbai 400085, India}
\affiliation{$^3$Electronic Structure Theory Group, Physical Chemistry Division, CSIR-National Chemical Laboratory, Pune 411008, India}
\affiliation{$^4$Indian Institute of Science Education and Research Kolkata, Mohanpur 741246, India}
%%%%%%%%%%%%%%%%%%%%%%%%%%%%%%%%%%%%%%%%%%%%%%%%%%%%%%%%%%%%%%%%%%%%%%%%%%%%%%%%%%%%%%%%%%%%%%%%%%%%%
\begin{abstract}
%The parity (${\mathcal{P}}$) and time-reversal (${\mathcal{T}}$) symmetry violating effects in paramagnetic heavy polar diatomic molecules are very important in 
%search of new physics beyond the standard model of particle physics. Therefore, 
We employ the $Z$-vector method %%and the expectation value approach 
in the four-component relativistic coupled-cluster framework to calculate the  parity (${\mathcal{P}}$) and time-reversal (${\mathcal{T}}$) symmetry violating scalar-pseudoscalar
(S-PS) nucleus-electron interaction constant ($W_\text{s}$), the effective electric field ($E_\text{eff}$) experienced by the
unpaired electron and the nuclear magnetic quadrupole moment (NMQM)-electron interaction constant ($W_\text{M}$) in the
open-shell ground electronic state of HgF. The molecular frame dipole moment and the magnetic hyperfine structure (HFS) constant 
of the molecule are also calculated at the same level of theory.
The outcome of our study is that HgF has a high value of $E_\text{eff}$ (115.9 GV/cm),
$W_\text{s}$ (266.4 kHz) and $W_\text{M}$ ($3.59\times 10^{33}$Hz/e.cm$^2$), which shows that it can be a possible candidate for the search of
new physics beyond the Standard model. %The uncertainty associated with our results is around 10\%.
Our results are in good agreement with the available literature values.
Furthermore, we investigate the effect of the basis set and that of the virtual energy functions on the computed properties. The role of the high-energy virtual spinors
are found to be significant in the calculation of the HFS constant and the ${\mathcal{P,T}}$-odd interaction coefficients.
\end{abstract}
%%%%%%%%%%%%%%%%%%%%%%%%%%%%%%%%%%%%%%%%%%%%%%%%%%%%%%%%%%%%%%%%%%%%%%%%%%%%%%%%%%%%%%%%%%%%%%%%%%%%%
%%%%%%%%%%%%%%%%%%%%%%%%%%%%%%%%%%%%%%%%%%%%%%%%%%%%%%%%%%%%%%%%%%%%%%%%%%%%%%%%%%%%%%%%%%%%%%%%%%%%%
\maketitle
%%%%%%%%%%%%%%%%%%%%%%%%%%%%%%%%%%%%%%%%%%%%%%%%%%%%%%%%%%%%%%%%%%%%%%%%%%%%%%%%%%%%%%%%%%%%%%%%%%%%%
\section{Introduction}
%%%%%%%%%%%%%%%%%%%%%%%%%%%%%%%%%%%%%%%%%%%%%%%%%%%%%%%%%%%%%%%%%%%%%%%%%%%%%%%%%%%%%%%%%%%%%%%%%%%%%
The matter-antimatter asymmetry in the universe is one of the biggest mysteries till date.  
The ongoing accelerator based experiments in the search for physics beyond the standard model (SM) of elementary particle physics can shed some light on
the dominance of matter over antimatter.
The violation of charge conjugation (${\mathcal{C}}$) and parity (${\mathcal{P}}$) is one of the several conditions that can explain the
matter-antimatter asymmetry \cite{sakharov}. However, the ${\mathcal{CP}}$ violation in Kobayashi-Maskawa model within the SM
is too weak to explain this asymmetry. Therefore, extra symmetry violating interactions that are missing in the Standard model
are necessary to explore the physics beyond the SM. Thus, the non-accelerator (i.e., low energy) experiments to study the ${\mathcal{P}}$ and time reversal
invariance (${\mathcal{T}}$) violation in nuclei, atoms and molecules become complementary to the accelerator-based high energy
experiments \cite{ginges_2004, sandars_1965, sandars_1967, labzovskii_1978, barkov_1980, shapiro_1968, pospelov_2005}.
Intrinsic electric dipole moment of electron (eEDM or $d_e$) \cite{bernreuther_1991, tl_edm, ybf_edm, titov_tho, tho_edm} and nucleon (proton or neutron)
\cite {engel_2013, pospelov_2005}, and
${\mathcal{P,T}}$-odd scalar-pseudoscalar (S-PS) nucleus-electron interaction \cite{tho_edm, sasmal_raf, kudashov_2014, sudip_hgh}
are the two major sources of the permanent electric dipole moment (EDM) of paramagnetic atomic and molecular systems.
% ${\mathcal{P,T}}$ violating interactions that give rise to
In addition, the ${\mathcal{P,T}}$-odd nuclear magnetic quadrupole moment (NMQM) interaction \cite{fleig_tan, flambaum_2017, titov_2014, dmitriev_1994} can
contribute to atomic and molecular EDM. Thus, the NMQM is an another way to study the nuclear ${\mathcal{P,T}}$-odd physics in an atom or molecule.
%one more way to carry the nuclear ${\mathcal{P,T}}$-odd interaction to an atom or molecule is through the
%nuclear magnetic quadrupole moment (NMQM) \cite{fleig_tan, flambaum_2017, titov_2014} 
%and sometimes, the NMQM induced by ${\mathcal{P,T}}$-odd inter nucleon interaction can have significant effect due to the nucleon EDM.
\par
According to the SM, the eEDM is too small ($< 10^{-38}$ e.cm) \cite{khriplovich_2011}
to be observed experimentally.
On the other hand, many extensions of the SM predict the value of eEDM to be in the range of $10^{-26}-10^{-29}$ e.cm \cite{commins_1999}.
The sensitivity of the modern eEDM experiment also lies in the same range.
Heavy polar diatomic paramagnetic molecules are suitable to be the candidate of modern eEDM experiments due to their high internal effective electric
field ($E_\mathrm{eff}$) \cite{sushkov_1978, flambaum_1976}.
Till date, the best upper bound limit of eEDM ($< 8.7 \times 10^{-29}$ e.cm) is obtained in the experiment carried out by the ACME collaboration \cite{tho_edm} using ThO.
In addition to the eEDM, the ${\mathcal{P,T}}$-odd S-PS nucleus-electron neutral current interaction, arising from the interaction between
pseudo-scalar electronic current and scalar hadronic current can also contribute to the ${\mathcal{P,T}}$-odd
frequency shift.
So, to interpret the results of experiment in terms of the eEDM and fundamental S-PS coupling
constant ($k_s$), we need to know the accurate value of $E_\mathrm{eff}$ and scalar-pseudoscalar ${\mathcal{P,T}}$-odd interaction constant ($W_\mathrm{s}$),
respectively. On the other hand,
the NMQM, one of the nuclear symmetry violating effects, may arise either due to the nucleon EDM or ${\mathcal{P,T}}$-odd nuclear forces. 
The nuclear EDM is often screened by the electrons in neutral atoms and molecules
and hence, their contribution is negligible to the measurable EDM of the system. However, the interaction of nuclear magnetic moment with
electrons is not screened.
% In this work, we are interested to study the NMQM produced by the ${\mathcal{P,T}}$-odd nuclear forces.
Also, the NMQM can produce larger EDM in paramagnetic systems than the Schiff moment \cite{flambaum_1984}. 
However, the NMQM effect only exists for the nuclei having nuclear spin $I>1/2$ in a paramagnetic system with non-zero electron angular momentum. 
The effect due to the NMQM increases significantly in deformed nuclei \cite{flambaum_1994}. Higher nuclear charge (Z) is another factor for the enhancement of the said effect.
The ${\mathcal{P,T}}$-odd interaction between the NMQM and the magnetic field produced by electrons in a paramagnetic system is significantly important as 
it can help in the investigation of new physics in the hadron sector of matter. 
Although, in recent times, the ${\mathcal{P,T}}$-odd  effects caused by the NMQM in various paramagnetic molecules have been studied,
further extensive study is important to unravel the new physics.
It is worth to mention that the accurate value of the NMQM interaction constant ($W_\mathrm{M}$) needs to be known
to interpret the experimentally measured ${\mathcal{P,T}}$-odd frequency shift due to the NMQM interaction in terms of the magnetic quadrupole moment ($M$) of nuclei.
However, the value of $E_\mathrm{eff}$, $W_\mathrm{s}$ and $W_\mathrm{M}$ 
cannot be experimentally measured and thus, we need to rely upon the accurate {\it ab initio} methods to calculate these quantities.
%{\color{red}
The `Atom in Compound' (AIC)\cite{titove_aic} properties such as $E_\mathrm{eff}$, $W_\mathrm{s}$ and $W_\mathrm{M}$ are very sensitive to the accuracy of the
wavefunction near the nuclear region and thus, an {\it ab initio} method which can efficiently incorporate both the relativistic and electron-correlation effects,
is necessary to calculate the AIC properties precisely. 
%The accuracy of these calculations can be estimated by comparing the theoretically calculated
%hyperfine structure (HFS) interaction constant with the available experimental values, because the HFS constant, like the ${\mathcal{P,T}}$-odd constants, also needs
%a precise wavefunction near the nuclear region.
%}  
\par 
%%%%%%%%%%%%%%%%%%%%%%%%%%%%%%%%%%%%%%%%%%%%%%%%%%%%%%%%%%%%%%%%%%%%%%%%%%%%%%%%%%%%%%%%%%%%%%%%%%%%
%In the electronic structure calculations, the effects of special relativity can be most elegantly treated via
%Dirac-Hartree-Fock (DHF) method in the four-component framework.
The relativistic effects  can be most elegantly treated using the four-component Dirac-Hartree-Fock (DHF) method in electronic structure theory. 
It approximates the electron-electron repulsion
in an average way which leads to the missing of correlation between opposite spin electrons. However, this missing
dynamic electron correlation can be efficiently incorporated in the single reference coupled-cluster (SRCC) method. Therefore, the relativistic
SRCC method is probably the most preferred many-body theory to deal with both the effects of correlation and
relativistic motion of electrons. Moreover, the properties of a many-electron system can be accurately calculated by the energy derivative
method in the SRCC framework. The $Z$-vector method \cite{schafer_1984, zvector_1989} is a widely used energy derivative technique for the calculation of
first order property. Recently, Sasmal {\it et al.} \cite{sasmal_pra_rapid} extended the $Z$-vector technique into the four-component relativistic 
coupled-cluster domain and successfully implemented the method to calculate various first order properties of atoms, ions and molecules.
%The $Z$-vector method can yield precise results in the SRCC framework. Therefore,

%%%%%%%%%%%%%%%%%%%%%%%%%%%%%%%%%%%%%%%%%%%%%%%%%%%%%%%%%%%%%%%%%%%%%%%%%%%%%%%%%%%%%%%%%%%%%%%%%%%%%
HgF is a highly polar and paramagnetic heavy
diatomic molecule. It has high diagonal Franck-Condon matrix element between the ground
and first excited electronic state, which is a characteristic of laser coolable molecule and the corresponding transition frequency
between these two states lies in the ultra visible region \cite{isaev_2010}. 
Recently, Vutha {\it et al.} \cite{vutha_2018} proposed an experiment in which
polar molecules such as HgF could be embedded in a solid matrix of inert gas atoms to measure the eEDM.
On the other hand,
Kozlov \cite{kozlov_1985, kozlov_1995} studied the ${\mathcal{P,T}}$-odd interactions in HgF using the relativistic semiempirical 
method. Similarly, Dmitriev {\it et al.} \cite{dmitriev_1992, kozlov_1995} performed quasirelativistic {\it ab initio} calculation of the aforementioned properties
using configuration interaction (CI) method. Das and co-workers \cite{prasanna_2015, prasanna_2017, abe_2018, sunaga_HgF} recently computed the $E_\text{eff}$ 
and $W_\text{s}$ in the HgF molecule using either the linear expectation-value or the finite field approach within the relativistic coupled-cluster model.
It is noteworthy that the calculations of ${\mathcal{P,T}}$-odd properties in heavy systems are usually difficult due to the strong interelectron correlations and the relativistic
motion of electrons, and the low level theoretical methods often do not produce accurate result. So, the precise calculations of 
these properties using the higher level of theory are always considered to be important.
For the calculation of first order properties, the energy derivative method such as $Z$-vector technique 
is usually more reliable than the expectation-value approach within the nonvariational SRCC framework. On the other hand, unlike the $Z$-vector method, the coupled-cluster 
amplitudes in the finite field approach depend on the external field parameters and the error associated with this approach is dependent on the number of data points
taken for the numerical differentiation. It is important to compute the aforementioned properties of HgF using a fully relativistic analytical (energy-derivative) method 
within the SRCC framework. Thus, the study of the AIC properties 
in HgF using more reliable {\it ab initio} method could be interesting. In this work, we have calculated the molecular dipole moment, the hyperfine structure
constant, $E_\text{eff}$, $W_\text{s}$ and $W_\text{M}$ of HgF in its ground electronic
($^2\Sigma$) state using the $Z$-vector method in the domain of four-component relativistic coupled-cluster theory. We also study the effect of the basis sets and that of the
virtual spinors on the computed properties. We believe that the present investigation of the effective electric field experienced
by the unpaired electron, the S-PS nucleus-electron
neutral current coupling and the nuclear magnetic quadrupole moment interaction with electrons in HgF
would be important in the search of new physics, and especially to examine the accuracy of the previously reported values in the literature.
%%%%%%%%%%%%%%%%%%%%%%%%%%%%%%%%%%%%%%%%%%%%%%%%%%%%%%%%%%%%%%%%%%%%%%%%%%%%%%%%%%%%%%%%%%%%%%%%%%%%%
The manuscript is organized as follows. The important aspects of the theory of the calculated properties and
the $Z$-vector approach in the domain of relativistic SRCC method are discussed in Sec. \ref{theory}.
Computational details are given in Sec. \ref{comp}. The results of the present work are presented and discussed 
in Sec. \ref{res_dis} and then, the concluding remark is given in Sec. \ref{conc}. 
Atomic units are used explicitly in the manuscript unless stated.
%Also, we have reported the absolute value of the calculated properties to avoid sign ambiguity. 
%%${\mathcal{P,T}}$-odd properties. 

%%%%%%%%%%%%%%%%%%%%%%%%%%%%%%%%%%%%%%%%%%%%%%%%%%%%%%%%%%%%%%%%%%%%%%%%%%%%%%%%%%%%%%%%%%%%%%%%%%%%%
\section{Theory}\label{theory}
%%%%%%%%%%%%%%%%%%%%%%%%%%%%%%%%%%%%%%%%%%%%%%%%%%%%%%%%%%%%%%%%%%%%%%%%%%%%%%%%%%%%%%%%%%%%%%%%%%%%%
\subsection{One-electron property operators}\label{prop}
%%%%%%%%%%%%%%%%%%%%%%%%%%%%%%%%%%%%%%%%%%%%%%%%%%%%%%%%%%%%%%%%%%%%%%%%%%%%%%%%%%%%%%%%%%%%%%%%%%%%%
The Hamiltonian for the interaction of the eEDM ($d_e$) with the internal molecular electric field \cite{kozlov_1987, titov_2006} is given as
\begin{eqnarray}
 H_d = 2icd_e \gamma^0 \gamma^5 {\bf \it p}^2 ,
\label{H_d}
\end{eqnarray}
where $\gamma$ are the usual Dirac matrices and {\bf \it p} is the momentum operator. Now, the $E_{\text{eff}}$ can be defined as
\begin{eqnarray}
 E_{\text{eff}} = |W_d \Omega|  = | \langle \Psi_{\Omega} | \sum_j^n \frac{H_d(j)}{d_e} | \Psi_{\Omega} \rangle |,
 \label{E_eff}
\end{eqnarray}
Here $\Omega$ is the projection of total electronic angular momentum on the molecular axis ($z$-axis). The value of $\Omega$ is 1/2 for the ground electronic 
state ($^{2}\Sigma$) of HgF. On the other hand, $\Psi_{\Omega}$ is the wavefunction of $\Omega$ state and $n$ is the total number of electrons. 

%%%%%%%%%%%%%%%%%%%%%%%%%%%%%%%%%%%%%%%%%%%%%%%%%%%%%%%%%%%%%%%%%%%%%%%%%%%%%%%%%%%%%%%%%%%%%%%%%%%%%
%%%                          Basis Information
%%%%%%%%%%%%%%%%%%%%%%%%%%%%%%%%%%%%%%%%%%%%%%%%%%%%%%%%%%%%%%%%%%%%%%%%%%%%%%%%%%%%%%%%%%%%%%%%%%%%%
\begin{table}[ht]
\caption{ Basis and cutoffs for virtual spinors used in our calculations.}
\begin{ruledtabular}
%%{%
\newcommand{\mc}[2]{\multicolumn{#1}{#2}}
\begin{center}
\begin{tabular}{lccccr}
\mc{4}{c}{Basis} & \mc{2}{c}{Virtual} \\
\cline{1-4} \cline{5-6} 
Name & Nature & Hg & F & Cutoff (a.u.) & Spinors \\
\hline
A & DZ & dyall.ae2z & cc-pVDZ & 50 & 153 \\
B & DZ & dyall.ae2z & cc-pVDZ & 1000 & 249 \\
%A & TZ & dyall.cv3z & cc-pCVTZ & -5000.0 & 1000.0 & 89 & 409 & -4.33244728 & -3.97959609 \\        %%22135.5 & 19257.0 \\
C & TZ & dyall.ae3z & cc-pVTZ & 50 & 261 \\  
D & TZ & dyall.ae3z & cc-pVTZ & 1000 & 427 \\      %%%%& 22135.5 & 19259.1 \\
%C & QZ & dyall.cv4z & cc-pCVQZ & -5000.0 & 50.0 & 89 &  &  &  \\
%C & QZ & dyall.ae4z & cc-pVQZ & -50.0 & 20.0 & 61 & 375 & -1.85948460 & -1.69271472 \\   %%% & 22296.6 & 19376.8 \\
E & QZ & dyall.ae4z & cc-pVQZ & 50 & 479 \\   %%%%& 22296.6 & 19686.8 \\
\end{tabular}
\end{center}
%}%
\end{ruledtabular}
\label{basis}
\end{table}
%%%%%%%%%%%%%%%%%%%%%%%%%%%%%%%%%%%%%%%%%%%%%%%%%%%%%%%%%%%%%%%%%%%%%%%%%%%%%%%%%%%%%%%%%%%%%%%%%%%%%
%%%%%%%%%%%%%%%%%%%%%%%%%%%%%%%%%%%%%%%%%%%%%%%%%%%%%%%%%%%%%%%%%%%%%%%%%%%%%%%%%%%%%%%%%%%%%%%%%%%%%
\begin{table}[ht]
\caption{Molecular frame dipole moment, $\mu$ (in Debye) and parallel component of magnetic HFS constant, A$_{\|}$ (in MHz) of HgF in different basis. Expt = Experiment.}
\begin{ruledtabular}
%{%
\newcommand{\mc}[1]{\multicolumn{#1}}
\begin{center}
\begin{tabular}{lccr}
Basis & $\mu$ & \mc{2}{c}{A$_{\|}$} \\
\cline{3-4}  %\cline{5-6}
 &  & $^{199}$Hg & $^{201}$Hg \\
\hline
A & 2.55 & 16795 & -6200 \\
B & 2.55 & 17256 & -6370 \\
%cv3z_1000 & 3.74 & 2.95 & 22135.5 & 19257.0 & & 8171.0 & 7108.5 & \\
C & 2.94 & 18730 & -6914 \\
D & 2.94 & 19259 & -7109 \\
%C & 3.17 & 19376.8 & 7152.7 \\
E & 3.16 & 19687 & -7267 \\
%Dmitrieve {\it et al.} &  & & 4.15 & 23310 & 24150 &  &  \\
Expt \cite{knight_1981} &  & 22621 & -8055 \\
\end{tabular}
\end{center}
%}%
\end{ruledtabular}
\label{hgf_hfs}
\end{table}

\begin{table*}[ht]
\caption{ Magnetic HFS constants (in MHz) of $^{199}$Hg$^{+}$, HgH and HgF. (Basis E is used for Hg$^+$ and HgF; dyall.cv4z basis for Hg and cc-pCVQZ for H, with 500 a.u. as cutoff for virtual spinors are used for HgH \cite{sudip_hgh}. Expt = Experiment.) }
\begin{ruledtabular}
%{%
\newcommand{\mc}[2]{\multicolumn{#1}{#2}}
\begin{center}
\begin{tabular}{lccccr}
Method & $^{199}$Hg$^{+}$ & \mc{2}{c}{$^{199}$Hg in HgH} & \mc{2}{c}{$^{199}$Hg in HgF}\\
\cline{3-4} \cline{5-6}
 & A$_J$ & A$_{\perp}$ & A$_{\|}$ & A$_{\perp}$ & A$_{\|}$ \\
\hline
Z-vector & 41423 & 6575 \cite{sudip_hgh} & 8440 \cite{sudip_hgh} & 19086 & 19687 \\    %%12441 (DHF), 8287.6 (CCSD)
%Z-vector & 41423 \footnote{basis: dyall.ae4z, cutoff for virtual spinors: 50 a.u.} & 6575 \footnote{basis: dyall.cv4z, cutoff for virtual spinors: 500 a.u.} \cite{sudip_hgh} & 8440 \footnote{basis: dyall.cv4z, cutoff for virtual spinors: 500 a.u.} \cite{sudip_hgh} & 19086 \footnote{basis: dyall.ae4z, cutoff for virtual spinors: 50 a.u.} & 19687 \footnote{basis: dyall.ae4z, cutoff for virtual spinors: 50 a.u.} \\    %%12441 (DHF), 8287.6 (CCSD)
Expt (in Ar) & 39600 \cite{knight_1972} & 6500(50) \cite{stowe_2002} & 8200(60) \cite{stowe_2002} & 21880(8) \cite{knight_1981} & 22621(10) \cite{knight_1981} \\
Expt (in Ne) & 41300 \cite{knight_1972} & 6200(3) \cite{stowe_2002} & 7780(5) \cite{stowe_2002} &  & \\
\end{tabular}
\end{center}
%}%
\end{ruledtabular}
\label{hfs_hg}
\end{table*}

%%%%%%%%%%%%%%%%%%%%%%%%%%%%%%%%%%%%%%%%%%%%%%%%%%%%%%%%%%%%%%%%%%%%%%%%%%%%%%%%%%%%%%%%%%%%%%%%%%%%%
\par
The interaction Hamiltonian for the S-PS nucleus-electron coupling \cite{hunter_1991} is defined as
\begin{eqnarray}
H_{\text{SP}}= i\frac{G_{F}}{\sqrt{2}}Zk_{s} \gamma^0 \gamma^5 \rho_N(r) ,
\label{H_SP}
\end{eqnarray}
where G$_F$ is the Fermi constant, and $\rho_N(r)$ is known as the nuclear charge density normalized to unity. The dimensionless constant
$k_s$ is defined as Z$k_s$=(Z$k_{s,p}$+N$k_{s,n}$), where N is the number of neutrons, and $k_{s,p}$ and $k_{s,n}$
are known as the electron-proton and electron-neutron coupling constant, respectively. The
$W_{\text{S}}$ can be evaluated from the following equation:
\begin{eqnarray}
 W_{\text{S}}=|\frac{1}{\Omega k_\text{s}}\langle \Psi_{\Omega}|\sum_j^n H_{\text{SP}}(j)| \Psi_{\Omega} \rangle|.
\label{W_s}
\end{eqnarray}
\par

The Hamiltonian for the NMQM-electron interaction \cite{flambaum_2017, kozlov_1987} is given by 
\begin{eqnarray}
H_{\text{MQM}}=-\frac{M}{2I(2I-1)}T_{ik}\frac{3}{2}\frac{[\vec{\alpha} \times \vec{r}]_i r_k}{r^5},
\label{H_MQM}
\end{eqnarray}
where {\boldmath $M$} is the NMQM tensor with components
\begin{eqnarray}
M_{ik}=\frac{3M}{2I(2I-1)}T_{ik} \\ 
T_{ik}=I_i I_k + I_k I_i-\frac{2}{3}\delta_{ik}I(I+1)
\end{eqnarray}
However, for the subspace of $\pm \Omega$, the equation \ref{H_MQM} can be written as \cite{flambaum_1984}
\begin{eqnarray}
H_{\text{MQM}}=-\frac{W_M M}{2I(2I-1)}\vec{S}^{\prime}\hat{T}\vec{n},
\end{eqnarray}
where $\vec{n}$ is the unit vector along the $z$-axis, and $\vec{S}^{\prime}$ is the effective electron spin. The
$W_M$ is defined as
\begin{eqnarray}
W_M=|\frac{3}{2 \Omega} \cdot \langle \Psi_{\Omega} | \sum_i^n \left( 
       \frac{\vec{\alpha}_i \times \vec{r}_i}{r_i^5} \right)_z r_z | \Psi_{\Omega} \rangle|.
 \label{W_M}
\end{eqnarray}
As the $^{199}$Hg nucleus has $I=1/2$, it cannot accomodate magnetic quadrupole moment. Thus, we focus on the $^{201}$Hg isotope ($I=3/2$) in HgF for the NMQM study.
%%%%%%%%%%%%%%%%%%%%%%%%%%%%%%%%%%%%%%%%%%%%%%%%%%%%%%%%%%%%%%%%%%%%%%%%%%%%%%%%%%%%%%%%%%%%%%%

We have mentioned earlier that accurate wave function near the nuclear region is necessary for the precise calculation of the above-mentioned matrix elements.
As the accurate calculation of the magnetic hyperfine structure (HFS) constant also depends on the accuracy of the core wave function, the
easiest way of testing the accuracy of the electronic wave function of interest is to compare the theoretically
calculated magnetic HFS constant with the experimentally measured value.

The matrix element for the magnetic HFS constant of the $J^{th}$ electronic state of an atom can be evaluated by the following expression:
\begin{eqnarray}
 A_J = \frac{\vec{\mu_k}}{IJ} \cdot \langle \Psi_J | \sum_i^n \left( 
       \frac{\vec{\alpha}_i \times \vec{r}_i}{r_i^3} \right) | \Psi_J \rangle,
 \label{hfs_atom}
\end{eqnarray}
where $\Psi_J$ is the wave function corresponding to the $J^{th}$ state, $I$ is known as the nuclear spin 
and $\vec{\mu}_k$ is nothing but the magnetic moment of the nucleus $k$.
Similarly, the parallel ($A_{\|}$) and perpendicular ($A_{\perp}$) components of the magnetic hyperfine structure constant
of a diatomic molecule can be defined as
\begin{eqnarray}
A_{\|(\perp)}= \frac{\vec{\mu_k}}{I\Omega} \cdot \langle \Psi_{\Omega} | \sum_i^n
\left( \frac{\vec{\alpha}_i \times \vec{r}_i}{r_i^3} \right)_{z(x/y)} | \Psi_{\Omega(-\Omega)}  \rangle,
\label{hfs_mol}
\end{eqnarray}

\par

%%%%%%%%%%%%%%%%%%%%%%%%%%%%%%%%%%%%%%%%%%%%%%%%%%%%%%%%%%%%%%%%%%%%%%%%%%%%%%%%%%%%%%%%%%%%%%%%%%%%%
\subsection{$Z$-vector method}\label{corr}
%%%%%%%%%%%%%%%%%%%%%%%%%%%%%%%%%%%%%%%%%%%%%%%%%%%%%%%%%%%%%%%%%%%%%%%%%%%%%%%%%%%%%%%%%%%%%%%%%%%%%
In a single reference theory, the Dirac-Hartree-Fock (DHF) is the most preferred method to take the relativistic
effects into account in an atom or molecule. It treats the inter-electronic repulsion in an average way and misses the correlation of the opposite
spin electrons. However, the dynamic part of the electron correlation needs to be incorporated by the SRCC method. 
So, the DHF wavefunction ($\Phi_0$) is used as a reference function for the proper treatment of the correlation effects in the SRCC method.
In the present work, we have used the four-component Dirac-Coulomb (DC) Hamiltonian which is defined as 
%\begin{eqnarray}
%H=\sum_i \Big [-c\vec{\alpha}_i\cdot\vec{\nabla}_i + (\beta -I)c^2 + V^{nuc}(r_i)
%+ \sum_{j>i} \frac{1}{r_{ij}} \Big ].
%\end{eqnarray}
\begin{eqnarray}
{H_{DC}} &=&\sum_{i} \Big [-c (\vec {\alpha}\cdot \vec {\nabla})_i + (\beta -{\mathbb{1}_4}) c^{2} + V^{nuc}(r_i)+ \nonumber\\
       && \sum_{j>i} \frac{1}{r_{ij}} {\mathbb{1}_4}\Big].%+\sum_{A,B>A} V_{AB},
\end{eqnarray}
Here, {\bf$\alpha$} and $\beta$ are known to be the Dirac matrices, $c$ is the speed of light,
${\mathbb{1}_4}$ is the 4$\times$4 identity matrix and $i$ denotes the electron.
$V^{nuc}(r_i)$ is the potential function for finite size nucleus, modelled by the Gaussian charge distribution.

The SRCC wavefunction has exponential ansatz which is given by
\begin{eqnarray}
|\Psi_{cc}\rangle=e^{T}|\Phi_0\rangle ,
\end{eqnarray}
$T$ in the above equation is known as the coupled-cluster excitation operator, which is defined
by
\begin{eqnarray}
 T=T_1+T_2+\dots +T_N=\sum_n^N T_n ,
\end{eqnarray}
with
\begin{eqnarray}
 T_m= \frac{1}{(m!)^2} \sum_{ij\dots ab \dots} t_{ij \dots}^{ab \dots}{a_a^{\dagger}a_b^{\dagger} \dots a_j a_i}.
\end{eqnarray}
The occupied (unoccupied) spinors are denoted by i,j(a,b) indices in the above expression. $t_{ij..}^{ab..}$ is the cluster amplitude corresponding 
to $T_m$ operator. In coupled-cluster singles and doubles (CCSD) model, $T=T_1+T_2$, and the unknown cluster amplitudes corresponding to T$_1$ and T$_2$ are solved 
using the following equations:
\begin{eqnarray}
 \langle \Phi_{i}^{a} | (H_Ne^T)_c | \Phi_0 \rangle = 0 , \,\,
  \langle \Phi_{ij}^{ab} | (H_Ne^T)_c | \Phi_0 \rangle = 0 ,
 \label{cc_amplitudes}
\end{eqnarray}
where, H$_N$ is the normal ordered DC Hamiltonian. Connectedness, which is denoted by the subscript $c$, ensures the size-extensivity.
By the term connectedness, we simply mean that only the connected terms survive in the
contraction between H$_N$ and T. Now, the correlated energy can be calculated by the following equation:
\begin{eqnarray}
E_{corr} = \langle \Phi_0 | (H_Ne^T)_c | \Phi_0 \rangle.
\end{eqnarray}

\par
%%%%%%%%%%%%%%%%%%%%%%%%%%%%%%%%%%%%%%%%%%%%%%%%%%%%%%%%%%%%%%%%%%%%%%%%%%%%%%%%%%%%%%%%%%%%%%%%%%%%%%%%
\begin{figure}[ht]
\centering
\begin{center}
\includegraphics[scale=.5, height=7cm]{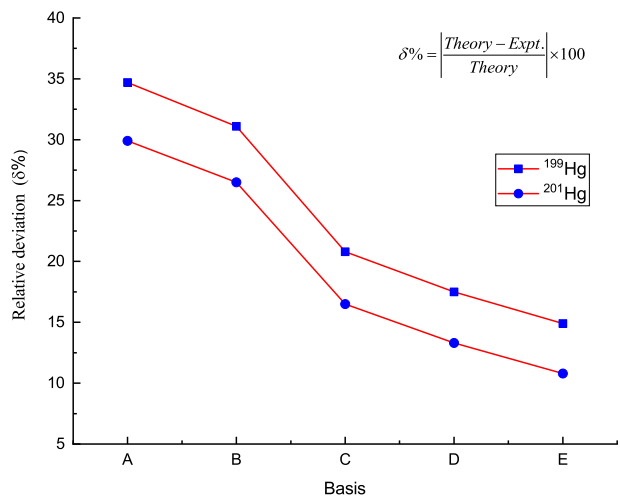}
\caption {Relative deviations of the parallel component of HFS constant of $^{199}$Hg and $^{201}$Hg in HgF at different basis.}
\label{deviation}
\end{center}  
\end{figure}
%%%%%%%%%%%%%%%%%%%%%%%%%%%%%%%%%%%%%%%%%%%%%%%%%%%%%%%%%%%%%%%%%%%%%%%%%%%%%%%%%%%%%%%%%%%%%%%%%%%%%
%%%%%%%%%%%%%%%%%%%%%%%%%%%%%%%%%%%%%%%%%%%%%%%%%%%%%%%%%%%%%%%%%%%%%%%%%%%%%%%%%%%%%%%%%%%%%%
%                     P,T-odd constants
%%%%%%%%%%%%%%%%%%%%%%%%%%%%%%%%%%%%%%%%%%%%%%%%%%%%%%%%%%%%%%%%%%%%%%%%%%%%%%%%%%%%%%%%%%%%%%
\begin{table}[ht]
\caption{W$_\mathrm{s}$ (in kHz), E$_\mathrm{eff}$ (in GV/cm), R (in 10$^{18}$/e.cm), and W$_\mathrm{M}$ (in 10$^{33}$Hz/e.cm$^2$) of HgF in different basis.}
\begin{ruledtabular}
%{%
%\newcommand{\mc}[3]{\multicolumn{#1}{#2}{#3}}
\begin{center}
\begin{tabular}{lccccr}
Basis & Nature & $W_\mathrm{s}$ & $E_\mathrm{eff}$ & R=$E_\mathrm{eff}$/W$_\mathrm{s}$ & $W_\mathrm{M}$\\
% & (kHz) & (GV/cm) & (10$^{18}$/e cm) & (10$^{33}$Hz/e cm$^2$) \\
%\cline{2-3} \cline{4-5} \cline{6-7}
%× & Expect. & Z-vector & Expect. & Z-vector & Expect. & Z-vector\\
\hline
A & DZ & 251.4 & 111.8 & 107.6 & 3.53 \\
B & DZ & 259.3 & 115.5 & 107.6 & 3.64 \\
C & TZ & 264.7 & 115.2 & 105.2 & 3.57 \\
D & TZ & 273.0 & 118.8 & 105.2 & 3.68\\
%C & QZ &  &  &  &  \\
E & QZ & 266.4 & 115.9 & 105.2 & 3.59 \\
\end{tabular}
\end{center}
%}%
\end{ruledtabular}
\label{hgf_pt}
\end{table}
%%%%%%%%%%%%%%%%%%%%%%%%%%%%%%%%%%%%%%%%%%%%%%%%%%%%%%%%%%%%%%%%%%%%%%%%%%%%%%%%%%%%%%%%%%%%%%
%%%%\
\par
%%%%%%%%%%%%%%%%%%%%%%%%%%%%%%%%%%%%%%%%%%%%%%%%%%%%%%%%%%%%%%%%%%%%%%%%%%%%%%%%%%%%%%%%%%%%%%%%%%%%%
%%%%%%%%%%%%%%%%%%%%%%%%%%%%%%%%%%%%%%%%%%%%%%%%%%%%%%%%%%%%%%%%%%%%%%%%%%%%%%%%%%%%%%%%%%%%%%
%                     Z-vector Energy Derivative
%%%%%%%%%%%%%%%%%%%%%%%%%%%%%%%%%%%%%%%%%%%%%%%%%%%%%%%%%%%%%%%%%%%%%%%%%%%%%%%%%%%%%%%%%%%%%%
%%%%%%%%%%%%%%%%%%%%%%%%%%%%%%%%%%%%%%%%%%%%%%%%%%%%%%%%%%%%%%%%%%%%%%%%%%%%%%%%%%%%%%%%%%%%%%
The coupled-cluster (CC) energy is a function of the molecular orbital coefficient (C$_M$) as well as the 
determinantal coefficient (C$_D$) in the expansion of the many electron correlated wavefunction. The CC 
equations are not usually solved variationally and thus the CC energy
is not optimized with respect to C$_M$ and C$_D$ for a fixed nuclear geometry \cite{monkhorst_1977}. 
Therefore, to calculate the CC energy derivative with respect to the perturbation, the derivatives of energy with respect to C$_M$ and C$_D$ along with the derivatives of
these two coefficients with respect to the external field of perturbation are required.
However, the derivative terms associated with C$_D$ can be included
introducing a perturbation independent linear operator $\Lambda$ \cite{zvector_1989}. It is an antisymmetrized
de-excitation operator. In the second quantized form, $\Lambda$ can be defined as 
\begin{eqnarray}
 \Lambda=\Lambda_1+\Lambda_2+ \dots+\Lambda_N=\sum_n^N \Lambda_n ,
\end{eqnarray}
where
\begin{eqnarray}
 \Lambda_m= \frac{1}{(m!)^2} \sum_{ij \dots ab \dots} \lambda_{ab \dots}^{ij \dots}{a_i^{\dagger}a_j^{\dagger} \dots a_b a_a} ,
\end{eqnarray}
Here $\lambda_{ab \dots}^{ij \dots}$ is the cluster amplitude corresponding  to $\Lambda_m$.
%The detailed description of $\Lambda$ operator and amplitude equation is given in Ref. \cite{zvector_1989}.
In CCSD framework, $\Lambda=\Lambda_1+\Lambda_2$. The explicit equations corresponding to the amplitudes of $\Lambda_1$
and $\Lambda_2$ can be written as
\begin{eqnarray}
\langle \Phi_0 |[\Lambda (H_Ne^T)_c]_c | \Phi_{i}^{a} \rangle + \langle \Phi_0 | (H_Ne^T)_c | \Phi_{i}^{a} \rangle = 0,
\end{eqnarray}
\begin{eqnarray}
% \begin{split}
\langle \Phi_0 |[\Lambda (H_Ne^T)_c]_c | \Phi_{ij}^{ab} \rangle + \langle \Phi_0 | (H_Ne^T)_c | \Phi_{ij}^{ab} \rangle \nonumber \\
 + \langle \Phi_0 | (H_Ne^T)_c | \Phi_{i}^{a} \rangle \langle \Phi_{i}^{a} | \Lambda | \Phi_{ij}^{ab} \rangle = 0.
% \end{split}
\label{lambda_2}
\end{eqnarray}
Once the amplitudes of the de-excitation operators are known, the energy derivative can be obtained by the following equation:
\begin{eqnarray}
 \Delta E' = \langle \Phi_0 | (O_Ne^T)_c | \Phi_0 \rangle + \langle \Phi_0 | [\Lambda (O_Ne^T)_c]_c | \Phi_0 \rangle
\end{eqnarray}
where, $O_N$ is known as the derivative of the normal ordered perturbed Hamiltonian with respect to the external field of perturbation.
%%%%%%%%%%%%%%%%%%%%%%%%%%%%%%%%%%%%%%%%%%%%%%%%%%%%%%%%%%%%%%%%%%%%%%%%%%%%%%%%%%%%%%%%%%%%%%%%%%%%%

%%%%%%%%%%%%%%%%%%%%%%%%%%%%%%%%%%%%%%%%%%%%%%%%%%%%%%%%%%%%%%%%%%%%%%%%%%%%%%%%%%%%%%%%%%%%%%%%%%%%%
\section{Computational details}\label{comp}
%%%%%%%%%%%%%%%%%%%%%%%%%%%%%%%%%%%%%%%%%%%%%%%%%%%%%%%%%%%%%%%%%%%%%%%%%%%%%%%%%%%%%%%%%%%%%%%%%%%%%
We have solved the DHF equation and generated the one-body, two-body and one-electron property integrals using the locally modified
version of DIRAC10 \cite{dirac10}. The finite size nuclear model with Gaussian charge distribution is considered, and the default
values of DIRAC10 is used as the nuclear parameters \cite{visscher_1997} in our calculations. Small and large component basis functions, which are scalar, are related to each other by the 
condition of restricted kinetic balance (RKB) \cite{dyall_2007}. The negative energy spectrum is removed in our calculations with the help of 
``no virtual pair approximation'' (NVPA). The properties of interest are calculated using the locally developed Z-vector code. 
The bond length of HgF is taken as 2.006 \AA \cite{bond_length} in all the calculations. 
We have used the following basis sets: double zeta: dyall.ae2z \cite{dyall_2010} for Hg, cc-pVDZ \cite{dunning_1989} for F, 
triple zeta (TZ) basis: dyall.ae3z \cite{dyall_2010} for Hg and cc-pVTZ \cite{dunning_1989} for F; quadruple zeta (QZ) basis: dyall.ae4z \cite{dyall_2012} basis for Hg 
and cc-pVQZ \cite{dunning_1989} basis for F. We consider all the occupied spinors in our calculations. However, we have excluded the virtual spinors having energy more than
a cutoff value. DZ basis with 50 and 1000 a.u. as cutoffs for virtual spinors are denoted as A and B, respectively. Similarly, TZ basis with 50 and 1000 a.u., and QZ with 50 a.u.
as virtual cutoffs are denoted as C, D, and E, respectively.
%None of the occupied spinors are frozen in our calculations. However, we have excluded the virtual spinors having energy more than a cutoff value. 
The details are given in table \ref{basis}.

%%%%%%%%%%%%%%%%%%%%%%%%%%%%%%%%%%%%%%%%%%%%%%%%%%%%%%%%%%%%%%%%%%%%%%%%%%%%%%%%%%%%%%%%%%%%%%%%%%%%%
\section{Results and discussion}\label{res_dis}
%%%%%%%%%%%%%%%%%%%%%%%%%%%%%%%%%%%%%%%%%%%%%%%%%%%%%%%%%%%%%%%%%%%%%%%%%%%%%%%%%%%%%%%%%%%%%%%%%%%%%
%%%%%%%%%%%%%%%%%%%%%%%%%%%%%%%%%%%%%%%%%%%%%%%%%%%%%%%%%%%%%%%%%%%%%%%%%%%%%%%%%%%%%%%%%%%%%%%%%%%%%
%%%                              Comparison
%%%%%%%%%%%%%%%%%%%%%%%%%%%%%%%%%%%%%%%%%%%%%%%%%%%%%%%%%%%%%%%%%%%%%%%%%%%%%%%%%%%%%%%%%%%%%%%%%%%%%
\begin{table}[ht]
\caption{Comparison of W$_\mathrm{s}$ (in kHz), E$_\mathrm{eff}$ (in GV/cm), R (in 10$^{18}$/e.cm), and W$_\mathrm{M}$ (in 10$^{33}$Hz/e.cm$^2$) of HgF.}
\begin{ruledtabular}
%{%
%\newcommand{\mc}[3]{\multicolumn{#1}{#2}{#3}}
\begin{center}
\begin{tabular}{lcccr}
Method  & $W_\mathrm{s}$ & $E_\mathrm{eff}$ & R=$E_\mathrm{eff}$/W$_\mathrm{s}$ & $W_\mathrm{M}$\\
% & (kHz) & (GV/cm) & (10$^{18}$/e cm) & (10$^{33}$Hz/e cm$^2$) \\
%\cline{2-3} \cline{4-5} \cline{6-7}
%× & Expect. & Z-vector & Expect. & Z-vector & Expect. & Z-vector\\
\hline
Z-vector CCSD (this work) & 266.4 & 115.9 & 105.2 & 3.59 \\
LECCSD \cite{prasanna_2015, prasanna_2017} & & 115.4 &  & \\
FFCCSD \cite{abe_2018} & & 116.4 &  & \\
LECCSD \cite{sunaga_HgF} & 264.7 & 114.9 &  & \\
CI \cite{dmitriev_1992, kozlov_1995} & 185.0 & 99.3 &  & 4.80 \\
non-relativistic MRCI \cite{meyer_2006} & & 68.0 & & \\
MRCI \cite{meyer_2008} & & 95.0 & & \\
Semiempirical \cite{kozlov_1985, kozlov_1995} & 191.0 & 99.0 & & 4.80\\
analytic \cite{dzuba_2011} & 204.5 &  & 112.5 & \\
\end{tabular}
\end{center}
%}%
\end{ruledtabular}
\label{comp_pt}
\end{table}
%%%%%%%%%%%%%%%%%%%%%%%%%%%%%%%%%%%%%%%%%%%%%%%%%%%%%%%%%%%%%%%%%%%%%%%%%%%%%%%%%%%%%%%%%%%%%%%%%%%%%
\par

%\begin{eqnarray}
% d_e + 5.56 \times 10^{-21} k_s = d_e^{expt}|_{\!_{k_s=0}},
%\label{relation}
%\end{eqnarray}
%where $d_e^{expt}|_{\!_{k_s=0}}$ is the eEDM limit derived from the experimentally measured P,T-odd frequency shift at the limit k$_s$ = 0.
%%%%%%%%%%%%%%%%%%%%%%%%%%%%%%%%%%%%%%%%%%%%%%%%%%%%%%%%%%%%%%%%%%%%%%%%%%%%%%%%%%%%%%%%%%%%%%%%%%%%%
%%%%%%%%%%%%%%%%%%%%%%%%%%%%%%%%%%%%%%%%%%%%%%%%%%%%%%%%%%%%%%%%%%%%%%%%%%%%%%%%%%%%%%%%%%%%%%%%%%%%%
\begin{table*}[ht]
\caption{ Effect of virtual spinors in our calculations. The reported values of A$_{\|}$ are for $^{199}$Hg and those of $W_\mathrm{M}$ are for $^{201}$Hg nuclei.}
\begin{ruledtabular}
   %{%
\newcommand{\mc}[2]{\multicolumn{#1}{#2}}
\begin{center}
\begin{tabular}{lccccccr}
\mc{2}{c}{Basis} & \mc{2}{c}{Virtual} & A$_{\|}$ & $W_\mathrm{s}$ & $E_\mathrm{eff}$ & $W_\mathrm{M}$\\
\cline{1-2} \cline{3-4} %\cline{9-10}
Hg & F & Cutoff (a.u.) & Spinors & (MHz) & (kHz) & (GV/cm) & (10$^{33}$Hz/e.cm$^2$) \\
\hline
dyall.ae2z & cc-pVDZ & 50 & 153 & 16795 & 251.4 & 111.8 & 3.53 \\
dyall.ae2z & cc-pVDZ & 1000 & 249 & 17256 & 259.3 & 115.4 & 3.64 \\
dyall.ae2z & cc-pVDZ & 3000 & 281 & 17313 & 260.2 & 115.8 & 3.65 \\
dyall.ae2z & cc-pVDZ & No cutoff & 365 & 17389 & 261.5 & 116.5 & 3.67 \\
%dyall.ae3z & TZ &  & 50.0 & 89 & 261 & -2.23364949 & 18730.4 & 264.7 & 115.2 & 3.57 \\
%dyall.ae3z & TZ &  & 100.0 & 89 & 333 & -3.18045042 & 18954.6 & 268.6 & 116.9 & 3.62 \\
%dyall.ae3z & TZ &  & 500.0 & 89 & 411 & -3.99790106 & 19190.4 & 272.36 & 118.56 & 3.66 \\
%dyall.ae3z & TZ &  & 1000.0 & 89 & 427 & -4.14807770 & 19259.1 & 272.98 & 118.79 & 3.68 \\
\end{tabular}
\end{center}
    %}%
\end{ruledtabular}
\label{core_effect}
\end{table*}

%%%%%%%%%%%%%%%%%%%%%%%%%%%%%%%%%%%%%%%%%%%%%%%%%%%%%%%%%%%%%%%%%%%%%%%%%%%%%%%%%%%%%%%%%%%%%%%%%%%%%
%The purpose of the present work is the precise calculation of the ${\mathcal{P}}$,${\mathcal{T}}$-odd interaction constants {\it viz.}, $E_\text{eff}$, $W_\text{s}$
%and $W_\text{M}$ of HgF using Z-vector method in relativistic coupled-cluster framework and to assess the candidature of the said molecule in 
%the search of new physics. Like HFS interaction, these ${\mathcal{P,T}}$-odd constants strongly
%depend on the wavefunction near the nuclear region of the heavy atom in the molecule. Thus, the accuracy of our calculations is
%estimated from the comparison of theoretically calculated HFS constants with the available experimental values. 
%We have presented the used basis, cut-off and correlation energies in table \ref{basis}. 

The parallel component of the HFS constant ($A_{\|}$) along with the molecular frame dipole moment ($\mu$)
of HgF are presented in table \ref{hgf_hfs}. We have also compared our results with the available experimental values, and
the relative deviations of the calculated HFS constants from the experimental values in different basis are shown in figure \ref{deviation}.
Magnitude of both the HFS constant and dipole moment increase with the use of a higher basis set. This is due to the fact that the addition of higher angular momentum basis
functions improves the configuration space. It is also observed that the HFS value increases with the addition of higher energy virtual spinors. But the same is
not true for the molecular dipole moment.
The most reliable values of $A_{\|}$ of $^{199}$Hg and $^{201}$Hg in HgF are found to be 19687 and -7267 MHz, respectively, which are calculated
using basis E (QZ, 50 a.u.). The value of $\mu$ obtained with the same basis is 3.16 Debye. However, it is observed from 
the table \ref{hgf_hfs} and figure \ref{deviation} that our most reliable HFS results deviate from the available experimental values \cite{knight_1981} by 10 - 14\%.
This could be because of two reasons. Firstly, we have not taken into account the Bohr-Weisskopf effect, Breit and QED interactions in our calculations.
The Bohr-Weisskopf effect is important for the accurate calculations of the HFS constants of heavy systems \cite{bw_prl_95}. Similarlly, the Breit and QED interactions could 
be important for the accurate computation of the HFS values. 
%Therefore, our calculated HFS values may be deviating significantly from the experimental values due to the exclusion of these effects.
Secondly, there could be some significant uncertainty in the experimental HFS values itself. For better understanding of the second reason,
we present the HFS constant of $^{199}$Hg in HgH and HgF along with $^{199}$Hg$^+$ in table \ref{hfs_hg}. 
The Z-vector value of the HFS constant for HgH are taken from Ref. \cite{sudip_hgh}.
From table \ref{hfs_hg}, it is clear that the exprimental magnetic HFS values of both $^{199}$Hg$^+$ \cite{knight_1972} and $^{199}$Hg in HgH \cite{stowe_2002}
differ significantly in Ar matrix compared to Ne matrix. From the previous study of various spectroscopic properties of various molecules, 
it has been seen that the Ne-matrices provide the most `gas-like environment' \cite{jacox_1987}.
This is also evident from our magnetic HFS calculation of $^{199}$Hg$^+$ as the theoretical value differ only 0.3\% from the experimental value in Ne matrix whereas
this deviation is 4.4\% for Ar matrix.
On the other hand, the experimental measurement of the HFS values of HgF has only been reported in Ar matrix in the year of 1981 \cite{knight_1981}. 
So, the uncertainties associated with the experimental values of gas-phase HFS constants of HgF are still ambiguous.
We, therefore, suggest that further experiment for more accurate measurement of the HFS values of HgF is needed to be carried out with the advanced experimental
techniques and especially in Ne matrix.
Nevertheless, we would like to comment that the HFS results for HgF produced by the Z-vector method in the present work are reasonably accurate.
This is because we have used the same basis (dyall.ae4z) and cutoff (50 a.u.) for the calculation of magnetic HFS value of $^{199}$Hg$^+$ and the 
deviation from the experiment in Ne matrix is negligible (0.3\%).
\par

The calculated ${\mathcal{P}}$,${\mathcal{T}}$-odd interaction constants along with the ratio ($R$) \cite{dzuba_2011} of $E_\text{eff}$ to $W_\text{s}$ of HgF 
are summarized in table \ref{hgf_pt}. The magnitude of $E_\text{eff}$, $W_\text{s}$ and $W_\text{M}$ increases due to the resultant effect of the addition of
% as we use higher basis sets. However, it is observed from the table \ref{hgf_pt} that the effect of 
higher angular momentum basis functions and that of higher energy virtual spinors.
% on the ${\mathcal{P,T}}$-odd properties seems to be additive. 
The configuration space improves as we go from A (DZ, 50 a.u.) to C (TZ, 50 a.u.) and then from C (TZ, 50 a.u.) to E (QZ, 50 a.u.) due to the addition of higher angular
momentum basis functions and thus, the magnitude of the calculated properties increases. Also, the same trend is observed when we go from
A (DZ, 50 a.u.) to B (DZ, 1000 a.u.) or C (TZ, 50 a.u.) to D (TZ, 1000 a.u.) due to the addition of higher virtual energy functions. But unlike the HFS constant, 
the magnitude of the ${\mathcal{P,T}}$-odd constants decreases as we go from D (TZ, 1000 a.u.) to
E (QZ, 50 a.u.) basis. This is probably because of the fact that for the ${\mathcal{P,T}}$-odd properties, the effect of higher energy virtual spinors is more 
prominent than that of the higher angular momentum basis functions, especially when we go
from TZ to QZ basis. (We discuss the effect of the virtual spinors on the calculated properties at the end of this section.) 
%one more interesting point to observe is that the trends of the ${\mathcal{P,T}}$-odd properties are similar with that in the HFS constant
%at various cutoff for the virtual spinors in a particular basis set. However, 
Nonetheless, as the relative deviation of HFS constants of HgF is lowest in the basis E, we consider the ${\mathcal{P,T}}$-odd constants calculated with this basis
as the most reliable results. Thus, our most reliable values of $E_\text{eff}$, $W_\text{s}$ and $W_\text{M}$ are found to be 115.9 GV/cm,
266.4 kHz and 3.59$\times 10^{33}$Hz/e.cm$^2$, respectively.
%Except the $W_\text{M}$, all the values presented in this work are for $^{199}$HgF. The $W_\text{M}$
%is for $^{201}$HgF as $^{201}$Hg nucleus has nuclear spin, $I>1/2$.  
%These values are quite high which clearly signifies that HgF could be a very good 
%player in the ${\mathcal{P}}$,${\mathcal{T}}$-odd frequency shift experiment.
%Especially, the large value of effective electric field experienced by the unpaired electron in HgF indicates that
%the experimental limit of a frequency due to EDM of an electron would improve in HgF. Furthermore,
%due to the large S-PS nucleus-electron interaction constant, 
%the S-PS neutral current interaction would significantly contribute
%to the permanent EDM of the molecule along with the eEDM.
Large magnitude of $E_\text{eff}$ and $W_\text{s}$ suggest that the experimental sensitivity of eEDM experiment would be very high for HgF.
On the other hand, the value of $W_\text{M}$
in $^{201}$HgF is probably the largest among all the possible candidates considered till date. Especially, the value is larger than that of ThF$^+$ \cite{flambaum_2014}, 
ThO \cite{titov_2014, flambaum_2014}, TaN \cite{fleig_tan, flambaum_2014},
HfF$^+$ \cite{flambaum_2017, flambaum_2014}, YbF \cite{flambaum_2014} and BaF \cite{flambaum_2014}. 
Although the electronic structure parameter for the NMQM interaction with electrons in $^{201}$HgF is quite large, the NMQM effect may not be significantly enhanced in this molecule 
since $^{201}$Hg is not a highly deformed nucleus. %However, the $^{201}$Hg is a stable isotope. 
Still, we believe that the nuclear ${\mathcal{P}}$,${\mathcal{T}}$-odd NMQM effect can enhance the EDM of the said molecule to some 
extent. On the other hand, we have already mentioned above that Vutha {\it et al.} \cite{vutha_2018} proposed a new experiment in which
HgF could be embedded in a solid matrix of inert gas atoms to measure the eEDM.
Thus, HgF can be an experimental candidate in search of new physics.

In table \ref{comp_pt}, our results are compared with the other available reported values in literature. 
Previously, Kozlov \cite{kozlov_1985, kozlov_1995}  
calculated the ${\mathcal{P}}$,${\mathcal{T}}$-violating interaction constants of HgF by relativistic semi-empirical method. 
They reported the value of $E_\text{eff}$, $W_\text{s}$ and $W_\text{M}$ to be 99 GV/cm, 191 kHz and 4.8$\times 10^{33}$Hz/e.cm$^2$,
respectively. The first {\it ab initio} calculation for symmetry violating interaction constants of HgF was
performed by Dmitriev {\it et al.} \cite{dmitriev_1992, kozlov_1995} who reported the value of $E_\text{eff}$, $W_\text{s}$ and $W_\text{M}$ as 99.3 GV/cm,
185 kHz and 4.8$\times 10^{33}$Hz/e.cm$^2$,
respectively. We have seen that their $E_\text{eff}$ and $W_\text{s}$ values are smaller than the values reported by our method but for $W_\text{M}$,
the trend is reversed. Overall, their results are in good agreement with our values.
Dmitriev {\it et al.} \cite{dmitriev_1992, kozlov_1995} used relativistic effective core potential (RECP) approach in the molecular SCF calculation part and then 
treated correlation effects of electrons via configuration interaction (CI) method by exciting only three outer electrons. This means, their calculations are not free from 
the error associated with the core-polarization effect. Further,
minimal atomic basis set for F, and only five relativistic valence orbitals 5d$_{3/2}$, 5d$_{1/2}$, 6s$_{1/2}$, 6p$_{1/2}$, and 6p$_{3/2}$ for Hg were used in their calculation.
%Their method also overestimated the spin-orbital mixing. 
In that work, it was claimed that the error in the calculations is around 20\%. However, further 
{\it ab initio} study shows that in these calculations a fortuitous cancellation of various effects occured \cite{titov_2006}.
On the other hand, we have employed an analytical approach called the $Z$-vector (energy derivative) method in the four-component relativistic CCSD
framework, which incorporates both the relativistic and correlation effects of electrons in an elegant way, to calculate the ${\mathcal{P}}$,${\mathcal{T}}$-odd
interaction constants using sufficiently large relativistic basis (QZ) and explicitly correlating all the electrons.
Also, Meyer {\it et al.} calculated $E_\text{eff}$ of HgF to be 68 \cite{meyer_2006} and 95 \cite{meyer_2008} GV/cm using the 
(60 core-electron) effective core potential (ECP)-based nonrelativistic and 
quasirelativistic multi-reference CI (MRCI) method, respectively. They used nonrelativistic and quasirelativistic pseudopotentials in their calculations.
Their non-relativistic value is quite small compared to the results reported by us. This is expected as the relativistic effects
are also important for the precise calculation of the AIC properties in heavy systems.
%It is worth mentioning here that although CI can treat electron correlation to all orders of the many-body perturbation theory,
%the truncated CI is not size-extensive unlike CC method. 
With the help of analytic expression, Dzuba {\it et al.} \cite{dzuba_2011} reported $W_\text{s}$ as 204.5 kHz.
Moreover, Das and co-workers recently used linear expectation-value (LECCSD) \cite{sunaga_HgF, prasanna_2015}
and finite-field (FFCCSD) \cite{abe_2018} relativistic coupled-cluster methodology to study the effective electric field and S-PS coupling constant in HgF.
They reported the value of $E_\text{eff}$ as 115.4 \cite{prasanna_2015, prasanna_2017} and 116.4 \cite{abe_2018} GV/cm using the double-zeta (DZ) level of basis in the LECCSD and
the FFCCSD method, respectively. Further, using the LECCSD method with triple-zeta level of basis, Das and coworkers obtained the value of $E_\text{eff}$ to be 
114.9 GV/cm and that of $W_\text{s}$ to be 264.7 kHz in the HgF molecule \cite{sunaga_HgF}.
As the LECCSD method does not include the nonlinear terms, it could miss some parts of the electron correlation effcts in the molecular calculation.
In nonvariational coupled-cluster framework, the expectation-value approach can be thought of as an approximation to the energy-derivative technique since the 
energy-derivative is the corresponding expectation-value plus some extra terms. On the other hand, the error associated with the numerical method such as the FFCCSD 
depends on the number of data points considered for the numerical differentiation. However, the same is not true for the analytical method such as $Z$-vector. 
Nonetheless, their values fortuitously show good agreement with our results. 
%In Ref. \cite{prasanna_2017}, the theoretical analysis of effective electric fields experienced by the unpaired electron
%in mercury monohalides were presented in detail.
%So, it is clearly seen that our values are in good agreement with most of the previously reported values in the literature.
%As
%the $Z$-vector method in the four-component relativistic CC framework, being a higher level of theory, can calculate the atomic and molecular properties with better precision,
%we argue that our present study verify the authenticity of the previously reported ${\mathcal{P}}$,${\mathcal{T}}$-odd properties.
%The other important ${\mathcal{P}}$,${\mathcal{T}}$-odd effect called nuclear MQM-electronic electromagnetic
%field interaction in HgF was studied by Kozlov using semi-empirical method and $W_\text{M}$ 
%was reported as 4.8 $10^{33}$Hz/e.cm$^2$ in his work. On the oher hand, this constant in our calculation is found to be 3.59 $10^{33}$Hz/e.cm$^2$, i.e., 
%our value is a little lower than that of Kozlov. As we have calculated the said constant by {\it ab initio} method such as Z-vector technique within relativistic CCSD framework 
%using sufficiently large basis (QZ, dyall.ae4z), our value is more reliable than that of Kozlov. 
On the other hand, the most reliable value of $R$ in HgF is found to be 105.2 in the unit of 10$^{18}$/e.cm in our study. Dzuba {\it et al.} suggested 
in Ref. \cite{dzuba_2011} that $R$ is very important to achieve the model independent limit of $d_e$ and $k_s$. They also argued that $R$
would have a particular value for a specific heavy nucleus irrespective of the diatom. Previously, Dzuba {\it et al.} \cite{dzuba_2011} and Sasmal {\it et al.} \cite{sudip_hgh}
reported the value of $R$ for Hg as 112.5 in HgF and 104.8 in HgH, respectively in the unit of 10$^{18}$/e.cm. 
So, it is clearly seen that our reported value of $R$ in HgF (i.e., 105.2$\times$10$^{18}$/e.cm) is reasonably in good agreement with the previously reported results in the literature.

%In other words, nice agreement of the value of $R$ with other reported values implies the reliability of our calculation.  
\par

%The uncertainty in our calculation may arises due to the following reasons:
%(i) higher order relativistic effects as we have not ignored the Breit/Gaunt interaction,
%(ii) higher order correlation effects,
%(iii) incompleteness of basis set, and
%(iv) cutoff used for the virtual orbitals.
However, we cannot deny the fact that we have not taken a number of effects (higher order correlation and relativistic effects etc.)
into account in our calculations, due to which there could be some errors
in our results. The ${\mathcal{P}}$,${\mathcal{T}}$-odd properties described here mainly depend on the 
valence electron density in the core region of the heavy atom and hence, these properties are not very sensitive to the retardation and magnetic
effects \cite{quiney_1998_tlf, lindroth_1989}. However, the error due to the exclusion of the higher order relativistic effects 
in the calculation of AIC properties is obtained as 0.5\%-1\% from the previous study \cite{prl_90, prl_01, prd_02}.
The error associated with the missing higher order correlation effects
% is present in our calculations. We 
can be estimated by comparing our values with the full configuration interaction (FCI) results. But
FCI calculation for HgF is too much expensive and thus, is not possible to perform in the present work. Nevertheless, we can comment from our experience
that the error due to the absence of higher order correlation effects is around 3.5\%. Incompleteness of basis set is another 
source of error in our results. This error can be assessed by comparing our results obtained using TZ and QZ basis.
While going from TZ to QZ (i.e. C to E) basis, the values of the ${\mathcal{P,T}}$odd constants are changed by around 0.6\%.
%(However, the HFS value is changed by around 5\%. Thus, HFS values are much more sensitive to the quality of the basis set compared to the ${\mathcal{P,T}}$odd constants.)
%That is, there is no valid reason that the corresponding error would exceed 1\%.
So, we can expect that the corresponding error would not exceed 1\%.
Moreover, to obtain the most reliable results, we have used the sufficiently large Dyall's relativistic all-electron (dyall.ae4z, QZ) basis, which contains
extra core-correlating functions for the proper treatment of core-polarization effect. But we have excluded higher virtual spinors with energy more than 50 a.u. in
the calculations and thus, there could be some error due to the restriction of correlation space. %%{\color {red} you should cite your previosus paper here} 
Especially, the inner-core (1$s$-3$d$) electrons
need very high energy virtual spinors for proper correlation \cite{core_jcp, core_pra, talukdar_hfs}. 
%%({\color {red} cite your previous paper and Skripnikov's paper JCP 145,214301 (2016), PRA 95, 022507 (2017}).
To decrease this type of error, we need to consider
the higher energy virtual spinors in our calculation which will be too expensive and is beyond the scope of our present study.
However, we have calculated the properties of interest using dyall.ae2z basis at various cutoffs for the virtual spinors to estimate the error.
The calculated results are summarised in table \ref{core_effect}. The contribution of the higher energy virtual spinors above 50 a.u. is found to be
3.4\% for the HFS constant, 3.9\% for the S-PS interaction coefficient, 4.0\% for the effective electric field and 3.8\% for the NMQM interaction constant.
This means, the high-energy virtual spinors are important for the precise calculation of the AIC properties. Nevertheless,
we can comment that the error associated with the restriction of correlation space is about 4\%. 
Considering all the possible errors, we expect that the uncertainty in our most reliable results is within 10\%.
%Nonetheless, we have employed the $Z$-vector (energy derivative) method in four-component relativistic CCSD
%framework, which incorporates both the relativistic and correlation effects of electrons in an elegant way, to calculate the ${\mathcal{P}}$,${\mathcal{T}}$-odd
%interaction constants using sufficiently large relativistic basis and explicitly correlating all the electrons. Thus, our values are 
%reasonably reliable. 

%The dominant matrix element for $E_\text{eff}$ arises from s$_{1/2}$ and p$_{1/2}$ while for $W_\text{s}$, it arises from
%s$_{1/2}$ and p$_{3/2}$.

%%%%%%%%%%%%%%%%%%%%%%%%%%%%%%%%%%%%%%%%%%%%%%%%%%%%%%%%%%%%%%%%%%%%%%%%%%%%%%%%%%%%%%%%%%%%%%%%%%%%%

%%%%%%%%%%%%%%%%%%%%%%%%%%%%%%%%%%%%%%%%%%%%%%%%%%%%%%%%%%%%%%%%%%%%%%%%%%%%%%%%%%%%%%%%%%%%%%%%%%%%%
\section{Conclusion}\label{conc}
We have employed the $Z$-vector method in the four-component relativistic coupled-cluster framework 
to calculate the symmetry violating interaction constants- E$_\mathrm{eff}$, W$_\mathrm{s}$ and W$_\mathrm{M}$ of HgF.
We have also calculated the molecular-frame dipole moment of the same molecule and the %parallel and perpendicular component of 
magnetic HFS constants of $^{199}$Hg \& $^{201}$Hg in HgF. 
The most reliable value of $E_\mathrm{eff}$, $W_\mathrm{s}$ and $W_\mathrm{M}$ obtained by us are 115.9 GV/cm,
266.4 kHz and 3.59$\times 10^{33}$Hz/e.cm$^2$, respectively. The estimated error in our reported values is within 10\%. However,
the previously reported theoretical results are in good agreement with our values.
The large values of the ${\mathcal{P,T}}$-odd interaction constants confirm the candidature
of HgF for ${\mathcal{P,T}}$-odd experiment in search of new physics. Further, it is clearly understood from our study that the high-energy virtual
functions play a significant role in the accurate calculation of the AIC properties in the HgF molecule.
%%%%%%%%%%%%%%%%%%%%%%%%%%%%%%%%%%%%%%%%%%%%%%%%%%%%%%%%%%%%%%%%%%%%%%%%%%%%%%%%%%%%%%%%%%%%%%%%%%%%%
\section*{Acknowledgement}
%%%%%%%%%%%%%%%%%%%%%%%%%%%%%%%%%%%%%%%%%%%%%%%%%%%%%%%%%%%%%%%%%%%%%%%%%%%%%%%%%%%%%%%%%%%%%%%%%%%%%
%{\color{red}
Authors acknowledge the resources of the Center of Excellence in Scientific Computing at CSIR-NCL. K.T. acknowledges the CSIR for 
the fellowship and Dr. Himadri Pathak and Dr. Sudip Sasmal for their insightful suggestions.
%}
%%%%%%%%%%%%%%%%%%%%%%%%%%%%%%%%%%%%%%%%%%%%%%%%%%%%%%%%%%%%%%%%%%%%%%%%%%%%%%%%%%%%%%%%%%%%%%%%%%%%%
%\bibliography{edm} % the name of your .bib file
%\bibliographystyle{h-physrev}  % the name of your .bst file

%%%%%%%%%%%%%%%%%%%%%%%%%%%%%%%%%%%%%%%%%%%%%%%%%%%%%%%%%%%%%%%%%%%%%%%%%%%%%%%%%%%%%%%%%%%%%%
\end{document}